%% file: PE140819.tex
\documentclass[aps,prl,twocolumn,superscriptaddress,reprint]{revtex4-1}

\usepackage{booktabs}
\usepackage{graphicx} 
\usepackage{color} 
\usepackage{dcolumn}  
\usepackage{bm}       
\usepackage{epstopdf}
\usepackage{amsmath,amssymb}

\input loc

\input dft

\def\hxc{_{\sss HXC}}

\def\sc{^{\rm sc}}
\def\LDAX{^{\rm LDAX}}

\begin{document}


\title{Almost Exact Exchange At Almost No Cost}

\author{Peter Elliott}
\affiliation{Max Planck Institute of Microstructure Physics, 
Weinberg 2, 06120 Halle (Saale), Germany}

\author{Attila Cangi}
\affiliation{Max Planck Institute of Microstructure Physics, 
Weinberg 2, 06120 Halle (Saale), Germany}

\author{Stefano Pittalis}
\affiliation{CNR-Istituto di Nanoscienze,
Via Campi 213A, I-41125 Modena, Italy}

\author{E.K.U. Gross}
\affiliation{Max Planck Institute of Microstructure Physics, 
Weinberg 2, 06120 Halle (Saale), Germany}

\author{Kieron Burke}
\affiliation{Department of Chemistry,
University of California, Irvine, CA 92697, USA}

\date{\today}

\begin{abstract}
A recently developed semiclassical approximation to 
exchange in one dimension is shown to be
almost exact, with essentially
no computational cost.  The variational
stability of this approximation is tested, and its far greater
accuracy relative to local density functional calculations demonstrated. 
Even a fully orbital-free potential-functional calculation 
(no orbitals of any kind) yields little error relative to
exact exchange, for more than one orbital.
\end{abstract}


\maketitle


Electronic structure problems in chemistry, physics, and materials science
are often solved via the Kohn-Sham  method
of density functional theory (DFT)\cite{HK64,KS65},
which balances accuracy with computational cost.
For any practical calculation, the exchange-correlation (XC) energy must be approximated
as a functional of the density.
The basic theorems of DFT guarantee its uniqueness, 
but give no hint about constructing approximations.
The early local density approximation (LDA)\cite{KS65}, much used 
in solid state physics, was the starting point for today's more accurate methods such 
as the generalized gradient\cite{B88,PBE96} and hybrid\cite{B93} approximations.
But a systematic approach for deriving these
has not yet been found, a fact that is reflected by the plethora
of XC approximations that are continuously created\cite{B12}.

This lack also inspires many approaches beyond
traditional DFT,
such as orbital-dependent functionals like exact exchange (EXX)
\cite{YW02,KK08}, use of
the random phase approximation\cite{EF11},  and
(first-order) density matrix functional theory\cite{DP78}.
While any of these can produce higher accuracy, their computational cost is
typically much greater, and none have yet yielded a universal improvement over
existing Kohn-Sham (KS) DFT.
Perhaps the most ubiquitous DFT method is that of hybrid functionals, which
replace some generalized gradient exchange with exact exchange.  Hybrids are now standard
in molecular calculations, and yield more accurate thermochemistry in most
cases\cite{B12}.  Furthermore, range-separated hybrids \cite{HSE03},
where the exchange
is treated in a Hartree-Fock fashion,
typically yield much improved band gaps
for many bulk solids\cite{HPSM05}.  
However, their computational cost in plane-wave codes
can be up to a thousand times higher
than that of generalized gradient approximation calculations,
making such methods much less useful in practice\cite{NTGV11}.
Implicit within quantum mechanics is that exact or approximate solutions of the Schrodinger equation are functionals of the {\em potential}, but the methodology of
DFT is built around the density instead.
Pioneering work\cite{YAW04} derived the duality of density and potential
functionals in the context of an orbital-dependent
KS-DFT calculation.
More recently, the formalism of a pure potential functional theory
(PFT) has been developed\cite{ELCB08,GP09,CLEB11}, and
approximations for non-interacting
fermions in simple model systems have been tested\cite{CLEB10,CGB13}.  
The leading corrections to Thomas-Fermi theory are explicit functionals
of the potential\cite{S80,S81,CLEB10}, and inclusion of these corrections yields approximations
that are typically much more accurate than their DFT 
counterparts.  

\begin{figure}[h]
\includegraphics[angle=0,width=8cm]{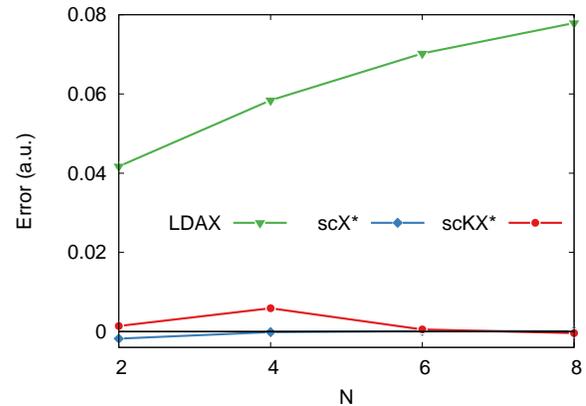}
\vskip -0.3cm
\caption{Error in the total energy made by
LDA exchange (LDAX),
semiclassical exchange (scX), 
and semiclassical kinetic and exchange (scKX) 
for $N$ spin-unpolarized, interacting fermions in a well
(see Tab.~\ref{t:D5m1.E.sc}).}
\label{f:dE_onXlda}
\end{figure}

Here we go beyond the non-interacting case by
including both Hartree and exchange components of
the electron-electron interaction to  illustrate the
promise of PFT for avoiding the cost of orbital-dependent
DFT calculations.  
In Fig.~\ref{f:dE_onXlda}, we show the errors made in the total energy 
of one-dimensional electrons in a potential with box boundaries,
using a recently developed semiclassical PFT approximation\cite{CEGB14}.
scX is a simple explicit 
formula for the exchange energy in terms of the KS potential, and here
is evaluated on the self-consistent potential 
in a  LDA exchange (LDAX) calculation.  
Even for only one occupied orbital, the error is less than 5\% of LDAX,
and is negligible for two or more orbitals, even though the exchange energy grows.
If such a formula existed for
three dimensions, the cost of (almost) EXX would be vanishingly small.
We also 
(i) develop the KS equation of PFT for interacting particles without recourse to DFT quantities, 
(ii) give an algorithm for solving this equation, 
(iii) implement that algorithm in 1d, and 
(iv) perform purely PFT calculations.


To begin, the ground-state energy of $N$ electrons in an external potential
$v(\br)$ is given by
\ben\label{gs.energy.trad}
E_0 = \min_\Psi \brk{\Psi}{\hat T + \hat V\ee + \hat V}{\Psi}\,,
\een
where the search is over all normalized, antisymmetric $\Psi$, and
$\hat T$ is the kinetic energy operator, 
$\hat V\ee$ the electron-electron repulsion,
and $\hat V= \sum_i v(\br_i)$ the one-body operator.
We use Hartree atomic units ($e^2=\hbar=m_e=1$) and suppress spin indices for
simplicity.
The universal potential functional\cite{CLEB11} is
\ben
F[v] =  \brk{\Psi_0[v]}{\hat T + \hat V\ee}{\Psi_0[v] }
\een
where $\Psi_0[v]$ is the ground-state wavefunction of $v(\br)$, so
\ben\label{search.v}
E_0 = \min_{\tilde v} \left(F[\tilde v] + \int d\br\ n[\tilde v](\br)\,v(\br) \right)
\een
where $\n[v](\br)$ is the ground-state density of $v(\br)$.
In the exact case, $\tilde v(\br)=v(\br)$, but this is not necessarily
true for approximations.

In previous work\cite{CLEB11}, it was shown that in PFT, once $n[v](\br)$ is
given, $F[v]$ can be deduced, either by a coupling-constant integral
or a virial relation.  When applied to non-interacting fermions,
an approximation $n\s[v\s](\br)$ yields an approximation
$T\s[v\s]$, where $v\s(\br)$ is the potential in this non-interacting case. 
Now we introduce a direct approximation to the
XC energy, $E\xc[v\s]$, as a functional of the KS potential, and
ask: How can these two approximations be used to find the ground-state
energy of interacting fermions?   This question differs from that of
deducing the KS equations in DFT, because here the approximation is
a potential functional, not a density (or orbital-dependent) functional.

To deduce the answer, we write the potential functional as a functional
of $v\s(\br)$ rather than $v(\br)$\cite{GP09}:
\ben
\bar F[v\s] = F[v[v\s]] = T\s[v\s] + U[v\s] + E\xc[v\s],
\een
i.e., all are functionals of the KS potential (which is uniquely determined
by $v(\br)$), where $U$ is the Hartree energy and $E\xc$ is everything else.
As mentioned above, with a given $n\s[v\s](\br)$, we can determine $T\s$
and $U$.
Applying \Eq{search.v}, but now searching over trial KS potentials,
yields, via the Hohenberg-Kohn theorem\cite{YAW04}
\ben\label{search.vs}
E_0 = \min_{v\s} \left(\bar F[v\s] + \int d\br\ n\s[v\s](\br)\,v(\br) \right)
\een
where we call the minimizing KS potential $\tilde v\s(\br)$. 
To find $\tilde v\s(\br)$, 
we use the Euler equation\cite{GP09}:
\ben
\left. \frac{\delta E_{v_0}[v\s]}{\delta v\s(\br)} \right|_{\tilde v\s} = 0
\een
for both the interacting and non-interacting systems, and equate potentials:
\ben\label{vs.self}
v'\s[\tilde v\s](\br) = v_0(\br) 
+ \int d\br'\ \chi\s^{-1}[\tilde v\s](\br',\br)\,
\left.\frac{\delta E\Hxc[v\s]}{\delta v\s(\br')}\right|_{\tilde v\s}\,,
\een
where $E\Hxc=U+E\xc$,
$\chi\s[\tilde v\s](\br',\br) = \left.\delta n\s[v\s](\br')/
\delta v\s(\br)\right|_{\tilde v\s}$
is the 
one-body density-density response function and:
\ben
v'\s[\tilde v\s](\br) = - \int \chi\s^{-1}[\tilde v\s]\left.\frac{\delta T\s[v\s]}{\delta v\s}\right|_{\tilde v\s} \,.
\een
as shown in Ref. \onlinecite{GP09}.
Note that
$v'\s = \tilde v\s$ only 
if $T\s[v\s]$ and $n[v\s]$ together satisfy the
noninteracting Euler equation, such as for the exact functionals, as in 
Ref. \cite{YAW04}.
The solution of \Eq{vs.self}
yields the minimizing KS potential $\tilde v\s(\br)$, 
once $n\s[v\s](\br)$ and $E\hxc[v\s]$ are given.
But since \Eq{vs.self} requires computing the inverse of
$\chi\s$, which becomes costly with increasing particle number,  
we instead directly minimize \Eq{search.vs}.

We next turn to actual calculations, using approximate potential functionals.
Applying similar integration techniques in the complex energy plane 
as in Refs.~\onlinecite{ELCB08} and \onlinecite{CLEB10},
we obtain a semiclassical potential functional approximation (PFA) 
to the one-body reduced density matrix 
\ben\label{gamma.sc}
\gamma\s^{sc}(x,x') = \sum\limits_{\lambda=+,-} 
\frac{\lambda\sin[\theta\F^\lambda(x,x')] \mathrm{cosec}{[\alpha\F^\lambda(x,x')/2] }}
{2T\F\sqrt{k\F(x)k\F(x')}}\,, 
\een
of $N$ fermions in a one-dimensional potential, inside a box, whose chemical potential is
above the potential everywhere. 
Here
$\theta^\pm(x,x') = \theta(x)\pm\theta(x')$, 
$\alpha^\pm(x,x')= \alpha(x)\pm\alpha(x')$,
$\theta(x) = \int_0^x dx'\ k(x')$ denotes the semiclassical phase,
$k(x)=\sqrt{2(\cE-v(x))}$ the wave vector, $\cE$ is the energy,
$\alpha(x) = \pi\tau(x)/T$, 
$\tau(x) = \int_0^x dx'\, k^{-1}(x')$ the traveling time
of a classical particle in the potential $v(x)$ from one boundary 
to the point $x$ at a given energy, and $T= \tau(L)$\cite{ELCB08}.
A subscript $F$ denotes evaluation at the Fermi energy, which
is found by requiring the wavefunctions to vanish at the edge,
i.e., $\Theta_F(L)=(N+1/2)\pi$.
The derivation and implications for DFT of
this expression is given elsewhere\cite{CEGB14}.
As $x\to x'$, the diagonal reduces to the
known semiclassical approximation for the
density\cite{ELCB08} and two derivatives yield the corresponding
approximation to the non-interacting kinetic energy\cite{CLEB11}.
For a given electron-electron repulsion, $v\ee(u)$, where 
$u=|x-x'|$ denotes the separation between electrons, 
the semiclassical exchange is:
\ben\label{Exsc}
E\x\sc[v\s] = -\half\int\limits_{-\infty}^\infty \int\limits_{-\infty}^\infty dx\,dx'\ 
|\gamma\s\sc[v\s](x,x')|^2\,v\ee(u).
\een


We next test both the accuracy and the stability of the semiclassical
approximations relative to standard DFT.
In all cases, we put the `electrons' in pairs in a 1d box of unit length,
with a one-body potential $v(x) = -5\sin^2(\pi x)$, and repelling each
other via $\exp(-\alpha u)$ with $\alpha=4$.
These parameters are chosen so that even for $N=2$, the condition
on the Fermi energy is satisfied.

We first define what \emph{exact} calculation 
we shall use to analyze our results. In this context,
it is a full OEP calculation using the
exact orbital expression for exchange.
Such a calculation produces the exact KS kinetic and exchange energies
and KS potential
on the self-consistent EXX density for the problem. 
Next, we define LDAX and check its performance.
The LDAX energy per electron is
\ben
\epsilon\x\LDA(n(x)) 
= -\frac{\arctan{\beta}}{\pi} 
+ \frac{\ln(1+\beta^2)}{2\pi\beta}
\een
with $\beta=2\pi n(x)/\alpha$.
In Tab.~\ref{t:D5m1.E.sc}
we report exact total energies and errors of several approximate
calculations, as a function of (double) occupation of orbitals.
We see that LDAX makes a substantial error for $N=2$ which grows
with $N$, although $E\x$ itself grows, so the fractional error is
vanishing (as it must\cite{ELCB08}) as $N\to\infty$.  
A modern generalized gradient approximation might reduce this error
by a factor of 2 or 3.
In Tab.~\ref{t:nrgs4}, we list the total energy and its various components 
for four particles in the well.  Since the energy error is almost
entirely given by the exchange error, this means the LDAX density and
component energies are very accurate, and the corresponding LDAX KS potential
quite accurate.  Due to the variational principle, the small differences
in the different energy components almost cancel.

Our first new calculation is a post-LDA calculation of the exchange energy
using the semiclassical approximation of Eq.~\ref{Exsc}, 
i.e., $E\sc\x[v\s\LDAX]$.  
This is orbital-free exchange but using the potential rather than the density
as the basic variable.
The error is plotted in Fig.~\ref{f:dE_onXlda}, and 
tabulated next to the LDAX results in Tabs.~\ref{t:D5m1.E.sc} and \ref{t:nrgs4}, 
denoted scX*, where the * indicates a non-variational calculation.  
Even for $N=2$, the error is an order of magnitude smaller
than LDAX.
As $N$ grows, the error shrinks very rapidly, even in absolute terms,
because the semiclassical corrections to LDAX capture the leading
corrections in powers of $1/N$\cite{CLEB10,CGB13}.  In the next column over, we even use
the semiclassical kinetic energy as well (scKX) on the LDAX KS potential,
and see that, although the errors can be much larger, they are still far
below those of LDAX.   These results show that the semiclassical exchange
and even kinetic energy can be extracted from a simple LDAX self-consistent
calculation, yielding much smaller errors than LDAX.
Thus results almost identical to expensive EXX OEP calculations can be found
at essentially no cost with a PFA exchange that includes the leading asymptotic
corrections to LDAX.

\begin{table}[htb]
\caption{Total EXX energy and respective errors of self-consistent 
as well as perturbative post-LDAX(*) calculations 
within LDAX, scX, and scKX for $N$ spin-unpolarized 
fermions interacting via $\exp(-4u)$ in an external potential $v(x) = -5\sin^2(\pi x)$ 
within a box of unit length.}
\label{t:D5m1.E.sc}
\begin{ruledtabular}
\begin{tabular}{ d  d d  d d d d d}
\multicolumn{1}{c}{$N$}              &
\multicolumn{1}{c}{$E^{\sss EXX}$} &  
\multicolumn{1}{c}{$E\x^{\sss EXX}$} &  
\multicolumn{5}{c}{error$\cdot 10^3$}\\
\hline
\multicolumn{3}{c}{}     & 
\multicolumn{1}{c}{LDAX}  &
\multicolumn{1}{c}{scX*}  &
\multicolumn{1}{c}{scKX*} &
\multicolumn{1}{c}{scX} &
\multicolumn{1}{c}{scKX}\\
\cline{4-6}
\cline{7-8}
2 &   2.81 & -0.52 & 41.72 & -1.79 &  1.40 & -3.10 & -29.60 \\
4 &  39.04 & -1.26 & 58.41 & -0.15 &  5.89 & -3.86 & - 1.14 \\
6 & 126.10 & -2.10 & 70.24 &  0.14 &  0.53 & -1.20 &   0.47 \\
8 & 283.70 & -2.98 & 77.91 &  0.08 & -0.40 & -0.10 & - 1.76 \\
\end{tabular}
\end{ruledtabular}
\end{table}
\begin{table}[htb]
\caption{Energy components of 
self-consistent calculations within LDAX, 
semiclassical exchange (scX), and a semiclassical approximation of all 
energy components (scKX) for $4$ 'electrons' in the
same problem as in Tab.~\ref{t:D5m1.E.sc}. }
\label{t:nrgs4}
\begin{ruledtabular}
\begin{tabular}{ d d d d d }
\multicolumn{1}{c}{} &
\multicolumn{1}{c}{EXX} &  
\multicolumn{3}{c}{error$\cdot 10^3$}\\
\hline
\multicolumn{2}{c}{} & 
\multicolumn{1}{c}{LDAX} &
\multicolumn{1}{c}{scX} &
\multicolumn{1}{c}{scKX}\\
\cline{3-5}
E     &  39.04 & 58.41  & -3.86  & -1.14\\
T\s   &  49.44 &  1.22  &  0.34  &  1.22\\
V\ext & -12.72 & -1.38  &  0.07  &  4.56\\
U     &   3.58 &  0.003 &  0.02  & -5.90\\
E\x   &  -1.26 & 58.56  & -4.29  & -1.02\\
\end{tabular}
\end{ruledtabular}
\end{table}

But such a recipe, while showing the accuracy of resulting exchange energies
quickly, can be criticized for not being variational, i.e., not the result of
any self-consistent minimization.  
Our second type of calculation is to again use the semiclassical PFT exchange
witihn a regular KS-DFT calculation. The resulting expression
for the total energy is then minimized.  We expand the KS potential
in Chebyshev polynomials 
and use the Nelder-Mead method\cite{NM65,PTVF92} to optimize
the expansion coefficients. A similar technique was
used for the EXX case\cite{CS99,PZCH12}, where the exchange energy
was the usual Fock integral. 
We should point out treating this method variationally required additional constraints than the perturbative case. For certain systems the minimization would find pathological potentials that behave badly near the box boundaries but nevertheless minimize the total energy. Conveniently the semiclassical approximations developed contain an error check in the form of the normalization of the semiclassical density. 
If this normalization deviated by $1\%$ or more from $N$,
we add a large penalty to the total energy. This is then used to exclude such potentials that lie far from the domain of applicability of our approximations and
 leads to the good results of Tab.~\ref{t:D5m1.E.sc}.

In Tabs.~\ref{t:D5m1.E.sc} and \ref{t:nrgs4}, next to the scKX* columns, we list the scX results of this
procedure.   The error remains much smaller than that of LDAX, and
rapidly reduces with increasing $N$.
This is consistent with our
previous semiclassical approximations for the density 
and kinetic energy\cite{ELCB08,CLEB10,CLEB11,CGB13}. 
However, errors are also typically much larger than those of the non-self-consistent
calculation (scX*), showing that the variational properties are less
robust than in LDAX.  This is not surprising, given that LDAX satisfies a
crucial symmetry condition that scX does not\cite{CLEB11,CGB13}.  This is related to
the very incorrect local minima that the procedure finds if not restrained,
as mentioned above.

\begin{figure}[t]
\centering
\includegraphics[width=0.5\textwidth]{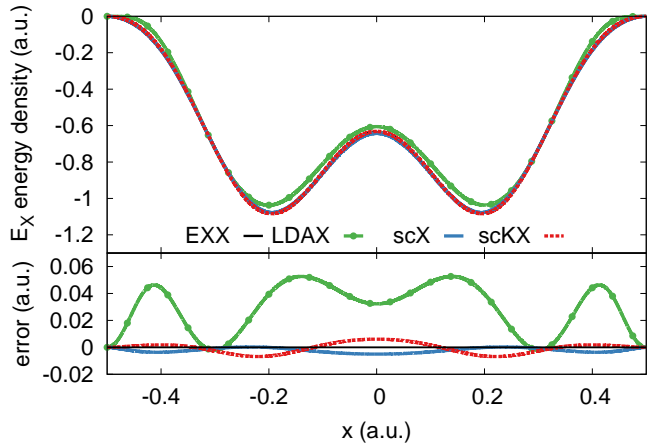}
\caption{Exchange energy density of 4 spin-unpolarized fermions
for the same problem as in Tab.~\ref{t:D5m1.E.sc}.  
The upper plot shows the EXX energy as well as result from a self-consistent calculation 
via LDAX, scX, and scKX. The respective errors are plotted in the lower panel.}
\label{f:exden.N2}
\end{figure}
To illusrate better the improvement in going from LDAX to scX, we plot the
exchange energy densities in
Fig.~\ref{f:exden.N2}, and their errors.
The scX density greatly improves over the LDAX density everywhere in space
(except where LDAX accidentally matches the exact value).  This is in stark
contrast to the well-known difficulty of defining and comparing energy
densities in generalized gradient approximations and other DFT approximations\cite{PRSB14}.

Finally, our {\em piece de resistance} is to run a pure PFT calculation, 
using semiclassical expressions for all energy components, not just the 
exchange energy, by directly minimizing \Eq{search.vs}.  This is a true
orbital-free calculation, the PFT analog of orbital-free DFT, and we
compare its results to a full OEP EXX calculation.  We denote this scKX,
and its results are in the far right columns of Tabs. \ref{t:D5m1.E.sc} and
~\ref{t:nrgs4}.

First, note that because we have now approximated the kinetic energy, we
would be doing extremely well to even match an LDAX calculation.   However,
we see that in every case, the errors are {\em smaller} than LDAX.  This
is the basic criterion for a successful orbital-free functional:  its errors
are smaller than typical errors in XC approximations.  However, we also note
that for any $N > 2$, its errors are so small (below 2 mH) that they
match those of exact exchange for most practical purposes. Finally, note that inaccuracies 
for $N=1$ or 2 do not matter, since the exchange energy for those cases is known exactly
via the Hartree energy. 

\begin{figure}[t]
\centering
\includegraphics[width=0.5\textwidth]{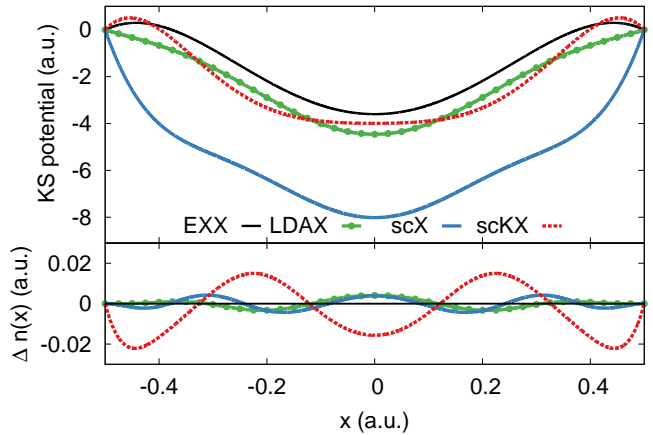}
\caption{Upper plot: Converged KS potentials of EXX, LDAX, scX, and scKX runs 
for the same problem as in Tab.~\ref{t:D5m1.E.sc} with 4 spin-unpolarized fermions.
Lower plot: Error in the respective, converged densities with respect to EXX.} 
\label{f:D5m1.vs.Dn.N2}
\end{figure}

Looking more closely, it is remarkable that scKX is more accurate
than scX for $N=2$ and $4$.  If we look at the individual energy
components in Tab.~\ref{t:nrgs4}, we see that, e.g.,  the Hartree energy is far more accurate in
scX than scKX, while the reverse is true for $E\x$.  This implies that
the density is quite inaccurate in scKX, but substantial cancellation of
errors occurs.  To see this,  in Fig.~\ref{f:D5m1.vs.Dn.N2} we plot both the KS potentials
and density errors for the different calculations, showing the much
greater errors in scKX.   However, the cancellation of errors might well
be due to the balanced nature of the calculation, since {\em all} energy
components have been derived from a single approximation for the density
matrix\cite{GP09}.  Only extensive testing for many different
circumstances can determine if this is a general phenomenon and if so,
where it fails.

%
%
%


Thus minimizing our PFA reproduces the result of a self-consistent EXX KS calculation. Furthermore, as the number of electrons increases, not only does the PFA computational effort not increase significantly, but the accuracy also increases. The Fock integral required in EXX or hybrid calculations scales formally as $\Omega^2N^2$ where $\Omega$ is the number of real space grid points used in our 1d box. Our semiclassical expression simply scales as $\Omega^2$. As $N$ increases, $\Omega$ should scale linearly in order to preserve the ratio of grid points to orbital nodes. Thus the Fock integral scales as $N^4$  while our approximation scales much more favourably as $N^2$. In quantum chemistry, evaluation of the Fock energy has been the focus of much effort to improve the scaling, but at best the scaling can be reduced to roughly $N^3$ (e.g., 
when localized basis sets and various optimization techniques are used). Thus our scaling remains advantageous. The scKX calculation is completely orbital-free and so avoids solving the KS equation. Either due to direct diagonalization or the orthogonalization of orbitals depending on the method used, the KS scheme scales as $N^3$, while scKX scales as $N^2$ due to exchange (the other energy components scale as $N$). Thus the PFT method can effectively reproduce the result of an EXX KS calculation while requiring a fraction of the computational cost.    
Substituting EXX with our semiclassical exchange may also be done in the hybrid functional approach to DFT (although treated within the OEP framework), where the fraction of EXX mixed in with a standard DFT functional may be replaced. Calculating this EXX energy is often the costliest part for hybrid calculations.
By using the semiclassical PFA exchange one could vastly speed up such calculations without a significant loss in accuracy.

In conclusion we have shown that 
an approximation to the exchange energy is \emph{almost}
exact and does not require any orbital information in the framework of PFT. 
In both accuracy and efficiency, the PFT method performs better than high-level KS-DFT calculations.
If ongoing work to extend the method to 3d systems is successful, 
electronic structure calculations could be sped up by serveral orders of magnitudes, 
allowing large systems that are currently out of reach with density functional methods 
to be studied.

PE, SP, and EKUG acknowledge funding by the European Commission (Grant No. FP7-NMP-CRONOS). EKUG thanks the KITP at UCSB for splendid hospitality.
This research was supported in part by the National Science Foundation under Grant No. NSF PHY11-25915. KB and AC acknowledge support by National Science Foundation under Grant No. CHE-1112442 NSF.

\bibliography{AC}

\end{document}

%% file: loc.tex
\def\Eq#1{Eq.~(\ref{#1})}

\def\n{n}

\def\sig#1#2{}

\def\bay{\begin{array}}
\def\eay{\end{array}}
\def\beit{\begin{itemize}}
\def\eit{\end{itemize}}

\def\bx{{\bf x}}

\def\F{_{\sss F}}

\def\cE{{\cal E}}

\def\crap#1{The following is crap:\\ #1}
\def\crap#1{\bf !!!!Erroneous entry was removed here!!!!}
\def\bei{\begin{itemize}}
\def\eei{\end{itemize}}
\def\benu{\begin{enumerate}}
\def\enu{\end{enumerate}}

\def\sc{^{\rm sc}}

\def\s{_{\rm s}}

\def\N{\tilde{N}}
\def\0{^{(0)}}

\def\1{^{(1)}}
\def\0{^{(0)}}

\def\2F1{_2\mathrm{F}_1}

\newcommand{\brk}[3]{\langle #1\,|\ #2 \ |\,#3 \rangle}

\def\br{{\bf r}}



%% file: dft.tex
\def\bea{\begin{eqnarray}}
\def\eea{\end{eqnarray}}
\def\ben{\begin{equation}}
\def\een{\end{equation}}
\def\benu{\begin{enumerate}}
\def\enu{\end{enumerate}}

\def\bei{\begin{itemize}}
\def\eei{\end{itemize}}
\def\beit{\begin{itemize}}
\def\eit{\end{itemize}}
\def\benu{\begin{enumerate}}
\def\enu{\end{enumerate}}

\def\n{n}

\def\sss{\scriptscriptstyle\rm}

\def\1var{(\bx_1...\bx\N)}

\def\half{\frac{1}{2}}

\def\br{{\bf r}}

\def\bx{{x}}

\def\x{_{\sss X}}

\def\s{_{\sss S}}
\def\xc{_{\sss XC}}

\def\Hxc{_{\sss HXC}}

\def\N{_{\sss N}}

\def\ext{_{\rm ext}}

\def\LDA{^{\rm LDA}}

\def\ee{_{\rm ee}}

\def\sph_int{ {\int d^3 r}}
